# Spin mixing conductance and spin magnetoresistance of iridate/manganite interface

G.A. Ovsyannikov[1]*, K.Y. Constantinian[1], V.A. Shmakov[1], A.L. Klimov[1,2], E.A. Kalachev[2], A.V. Shadrin[1,3], N.V. Andreev[4], F.O. Milovich[4], A.P. Orlov[1], P.V. Lega[1]

[1] Kotelnikov Institute of Radio Engineering and Electronics Russian Academy of Sciences Moscow125009, Mokhovaya 11-7, Russia.

[2] Russian University of Technology – MIREA, Moscow 119454, Vernadsky Av. 78, Russia.

[3] Moscow Institute of Physics and Technology (National Research University), Dolgoprudny 141701, Moscow region, Russia

[4] National Research Technological University - MISIS, Moscow 119049, Leninsky Prospekt 4, Russia.

*gena@hitech.cplire.ru

We present results on experimental studies of spin current, measured under spin pumping at ferromagnetic resonance in wide frequency band 2-20 GHz for $SrIrO_3/La_{0.7}Sr_{0.3}MnO_3$ heterostructures fabricated by RF magnetron sputtering at high temperature. The epitaxial growth of the thin film in heterostructure by a cube-on-cube mechanism was confirmed by XRD and TEM analysis. Taking into account the contribution of anisotropic magnetoresistance the spin current was estimated as 1/3 of the total response. We show that both real and imaginary parts of spin mixing conductance are valuable for heterostructures with strong spin-orbit interaction in $SrIrO_3$. Imaginary part of spin mixing conductance was estimated by means of shift of ferromagnetic resonance field of $La_{0.7}Sr_{0.3}MnO_3$ layer in heterostructure. The spin magnetoresistance was evaluated from angular dependencies of magnetoresistance measured in planar Hall configuration. In order to extract the influence of anisotropic magnetoresistance a $La_{0.7}Sr_{0.3}MnO_3$ film was measured as well. The spin Hall angle for heterostructure was found higher than for interface Pt/ $La_{0.7}Sr_{0.3}MnO_3$.

Keywords: heterostructures, manganite, iridate, ferromagnetic resonance, spin current, inverse spin Hall effect

## 1. Introduction

The use of spins of electron instead of charges opens up new opportunities in microelectronics, especially for reduction of heat dissipation in submicron-sized elements. The magnetization forced by microwaves at ferromagnetic resonance (FMR) generates pure (without charge transfer) spin current at the interface of metal/ferromagnet. It can be converted to charge current by means by inverse spin Hall effect

(ISHE) [1-8] . The generation of spin current and conversion of spin current into a charge current require a completely different approach in comparison to charge transfer electronics. A challenging task is the enhancement of efficiency of spin to charge conversion. The both spin Hall effect (SHE) and ISHE lead to a change of magnetoresistance in metal/ferromagnet (N/F) heterostructure depending on the magnetization direction, called spin Hall magnetoresistance (SMR) [9-12], accepted as an effective tool for probing the spin Hall angle and spin diffusion length.

The most common method is to use spin pumping in a presence of ferromagnetic resonance (FMR) at N/F interface. Indeed, under FMR a precessing magnetization in a ferromagnet generates spin current via spin pumping, which can be converted at the interface with an adjacent normal layer, to a dc voltage by ISHE. The amplitude of spin current depends on precessing magnetization and spin mixing conductance characterized by its real and imaginary parts. The conversion efficiency of spin current to the charge current is characterized by the spin-mixing conductance and the spin Hall angle $\theta_{SH}$. The last one could be evaluated as the ratio of the spin Hall magnetoresistance and the electro-resistance of the N metal by means of magnetotransport measurements [13-15].

Studies on spin pumping and induced by ISHE the dc voltages in a F/N heterostructure were first carried out with the Pt as a normal metal in combination with the permalloy (NiFe) as a F metal [4, 16, 17], and for structures using N metal contacting to insulating ferromagnetic yttrium iron garnet (YIG) [18-20]. At the same time a spin-orbit interaction plays rather a decisive role and a variety of metals with strong spin-orbit interaction have been used in combination with the metallic ferromagnet (NiFe) [21, 22].

Complex oxides displaying intriguing interplay between charge, spin, orbital and lattice degrees of freedom offer rich platform for both fundamental and application-oriented research due to their physical properties expanded from high temperature superconductivity in cuprates, colossal magnetoresistance in doped manganites [23], and an exotic band structure effects in perovskite iridates [24]. Moreover, the sensitivity of complex oxides to epitaxial strain [25] which influences the interface chemistry and crystal orientation provides opportunities for tuning the electronic and magnetic structure, leading also to spin-orbit interaction (SOI) effects. In particular, 5d transition metal oxides (TMO) with strong SOI and electron–electron correlation pushed on studies of nontrivial quantum phases [26–28], magnetic anisotropy manipulation [29] and intrinsic charge-spin interconversion [30–33]. However, despite of rich literature there are known just very few studies on charge-spin interconversion in 5d TMOs in epitaxial heterostructures. In these studies the estimated values of efficiency of spin current to charge current conversion were comparable to the N/F heterostructure.

In this paper we employ atomically matched interface of oxide ferromagnet and SOI TMO exhibiting metallic conductance, and coupled well chemically for comprehensive studies of charge-to-spin conversion processes. For our studies we choose the perovskites: manganite $La_{0.7}Sr_{0.3}MnO_3$ (LSMO) and the iridate SrIrO$_3$ (SIO). However, there are too many parameters, determining the magnitude of the spin current cased by spin pumping at FMR, which cannot be experimentally estimated with the good enough accuracy. As a result there is a wide scatter of experimentally obtained values of θ$_{SH}$ for the same heterostructure. At the same time, the number of parameters in the relationship between θ$_{SH}$ and SMR is much smaller.

The defining parameters for ISHE and SMR are the spin diffusion length in the N metal ($\lambda_{sd}$) the spin Hall angle (θ$_{SH}$) which quantifies the efficiency of spin to charge current conversion, and the spin mixing conductance ($G^{\uparrow\downarrow}$), which depends on the scattering matrices for electrons at the N/F interface and can be seen as the transparency of the interface for transfer of spin angular momentum [34]. The evaluation of the three above mentioned parameters is a delicate and the main task of this work.

The paper is organized as follows, besides the Introduction in Section I, results of fabrication and structural study of SrIrO$_3$/La$_{0.7}$Sr$_{0.3}$MnO$_3$ heterostructures are given in Section II with the details on the samples characterization and experimental setup. In Sec. III we discuss the experimental data on the voltage caused by charge current in spin pumping regime obtained for La$_{0.7}$Sr$_{0.3}$MnO$_3$ film and SrIrO$_3$/La$_{0.7}$Sr$_{0.3}$MnO$_3$ heterostructure. The real and imaginary parts of spin mixing conductance of the interface were determined from frequency dependence of FMR spectrum. In Sec. IV we discuss results of the magnetoresistance measurements both LSMO films and SrIrO$_3$/La$_{0.7}$Sr$_{0.3}$MnO$_3$ heterostructure. The value of the spin Hall angle was determined from angular dependence of longitudinal and transverse magnetoresistance. An influence of anisotropic magnetoresistance on the spin Hall magnetoresistance was observed. Conclusions are given in Sec. V.

**2. Manganite/iridate heterostructures**

Thin epitaxial films of strontium iridate SrIrO$_3$ (SIO) and manganite La$_{0.7}$Sr$_{0.3}$MnO$_3$ (LSMO) of nanometer thickness were grown on single-crystal substrates (110)NdGaO$_3$ (NGO) by magnetron RF sputtering at substrate temperatures of 770-800˚C in Ar and O$_2$ gas mixture at the total pressure of 0.3 mBar [35-37].

The crystal structure of the obtained heterostructures SIO/LSMO has been studied by X-ray diffraction analysis and transmission electron microscopy (TEM). The crystal lattice of SIO and LSMO could be described as a distorted pseudo-cube with lattice parameters $a_{SIO}$ = 0. 396 nm and $a_{LSMO}$ = 0.389 nm, respectively [36]. Figure 1 shows X-ray Bragg diffractogram of SIO/LSMO heterostructure with thick film of Pt as protected layer. Multiple reflections from plane (001) of SIO film and reflections (110)NdGaO$_3$ substrate, coinciding with the reflections from plane (001) of LSMO, as well as the reflections from Pt film are seen. A thick platinum film was deposited on top of heterostructure to avoid charge build-up. XRD data in Fig.1 allows us to conclude that the growth of heterostructure is performed by the "cube-on-cube" mechanism with the following ratios: (001)SrIrO$_3$||(001)La$_{0.7}$Sr$_{0.3}$MnO$_3$||(110)NdGaO$_3$ and [100]SrIrO$_3$||[100]La$_{0.7}$Sr$_{0.3}$MnO$_3$||[001]NdGaO$_3$ [36].

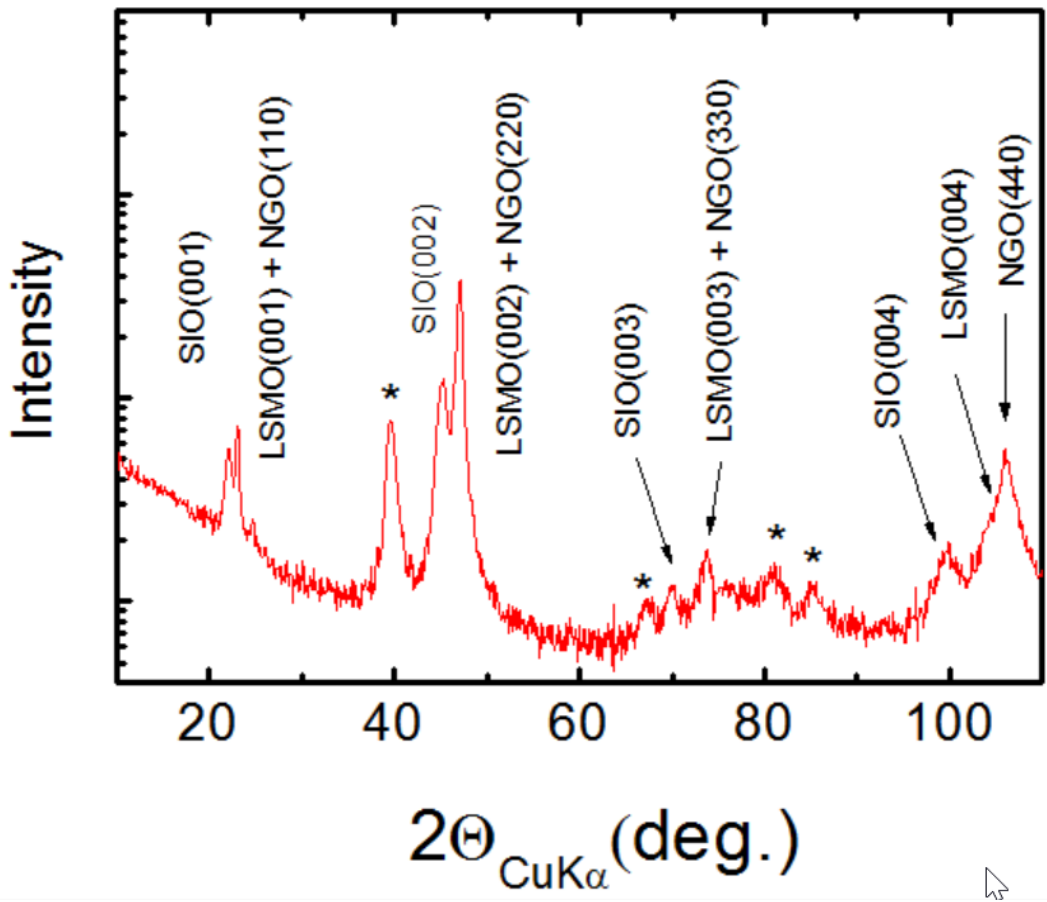

Fig.1 XRD Bragg reflections for Pt/SIO/LSMO/NGO heterostructure. The reflections from the Pt are marked by an asterisk.

Figure 2 shows a TEM image of a cross section of a heterostructure obtained with a transmission electron microscope JEM − 2100 at 200 kV. Elemental analysis was performed by X-ray energy dispersive system (OXFORD Instruments, INCA Energy). The cross section slice plate for transmission electron microscopy was made by using a focused ion beam in a Carl Zeiss CrossBeam Neon 40 EB scanning electron-ion microscope equipped with an auto-emission electron and a gallium ion gun with a resolution of 1 nm. The unit was equipped with a micromanipulator. To protect from damage layer of metal mask (Pt) was deposited by DC sputtering (100 nm) and then additionally up to 2 μm thick formed on the sample surface by the gas injection system for local precursor deposition. Ga+ ions with energy of 30 keV were used to obtain the slice and its thinning (polishing) with a gradual decrease of the etching current from 5 nA to 5 pA. At the final stage the ion energy was decreased to 5 keV to remove the broken at 30 keV layer.

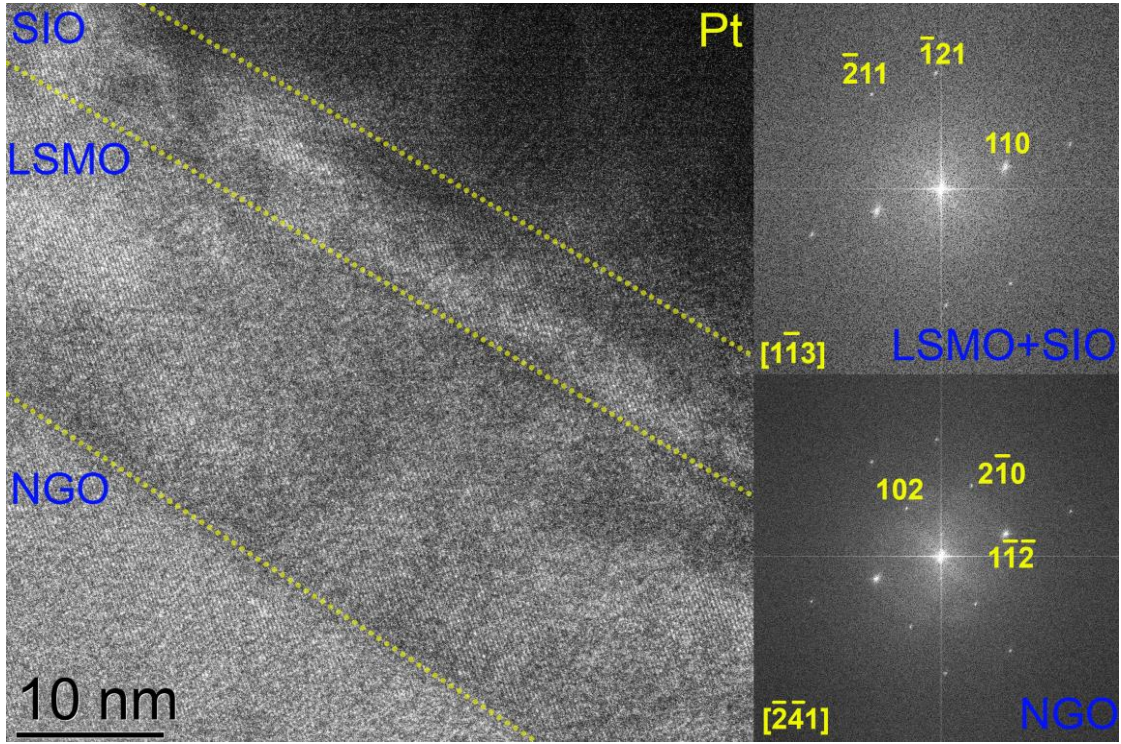

Fig.2. TEM image of a cross section of Pt/SIO/LSMO/NGO heterostructure covered with a thick layer of Pt playing the charge-streaming cladding role. The electron diffraction from the NGO substrate regions and LSMO/SIO films are shown on the right.

In the high-resolution image, we can observe a clearly pronounced interface between the LSMO layer and NGO substrate, as well between the SIO and LSMO layers. The right side of the figure shows Fourier images from NGO substrate and SIO/LSMO heterostructure layers. The coincidence of the reflexes in the Fourier image in the SIO/LSMO heterostructure layers indicates the proximity of the crystal lattices of the layers and the epitaxial growth of SIO on LSMO. The reflexes of the heterostructure also lie on part of the reflexes from the substrate (211)LSMO∥(204)NGO and (110)LSMO∥(112)NGO, which is in agreement with the assumption of epitaxial growth of the heterostructure on the substrate by the proposed cube-on-cube mechanism.

### 3. Spin current

In SIO/LSMO heterostructures paramagnetic conducting material SIO with a pronounced SOI used as a normal metal, while ferromagnet LSMO is a magnetic half-metal. The sample has shape of a strip deposited on NGO substrate with metal (Pt or Ag) contacts at the edges. For generation of spin current $j_S$ under FMR pumping SIO/LSMO heterostructures was placed either in a rectangular microwave cavity operating at $TE_{012}$ resonant mode at frequency F=9.0 GHz [36], or on the wide band microstrip line operating at F=2-20 GHz [37]. For the detection of spin current by ISHE the magnetic field voltage depend-ence V(H) was recorded by sweeping the external field H across the resonance value $H_0$, using data accumulation technique.

The typical signals V(H) detected at the SIO/LSMO heterostructure are shown in Fig. 3 for two frequencies F=2.6 GHz and F=9.0 GHz [35, 36, 38]. The experimental dependence could be approximated by sum of functions which take into account the effect of spin current generation under FMR pumping and the contribution from anisotropic magnetoresistance (AMR) [16, 39]:

$$V = [V_{AMR}^S L(H) + V_{AMR}^A D(H)] \sin 2\phi_0 \sin\phi_0 + V_Q L(H) \cos\phi_0 \tag{1}$$

where $L(H)=\Delta H^2/[(H-H_0)^2 + \Delta H^2]$ is the symmetric Lorentz function, $D(H)=\Delta H (H-H_0)/[(H-H_0)^2 + \Delta H^2]$ is the dispersion function, $V_{AMR}^S$ and $V_{AMR}^A$ are the amplitudes of symmetric and asymmetric parts of AMR, $V_Q(H)$ is the voltage caused by spin current $j_S$ flow through the SIO/LSMO interface, $\phi_0$ is the angle between dc magnetic field and the normal to voltage direction caused by ISHE. The ratio $V_{AMR}^A/V_{AMR}^S = -\text{tg}\phi_I$ [16] where $\phi_I$ is the phase difference between microwave current and microwave magnetization was obtained from the voltage dependence upon the angle $\phi_0$ [36]. For the case of asymmetric AMR $V_{AMR}^A = 0$ at $H=H_0=1957$ Oe (see Fig. 3b) we estimate the ratio $V_Q/V_{AMR}^S = 0.3 \pm 0.03$ The experimental value of $V_Q$= 1.7 µV was got at the maximal power of microwave pumping by Gunn diode at F=9 GHz.

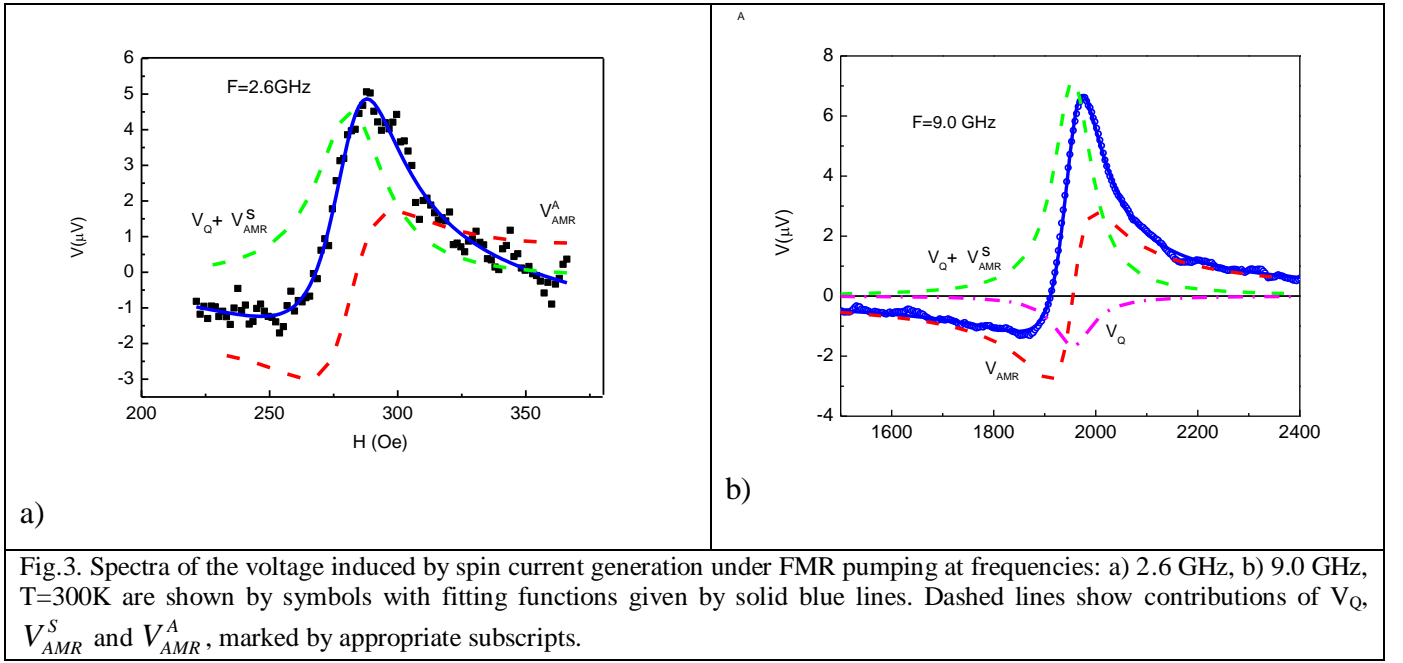

Fig.3. Spectra of the voltage induced by spin current generation under FMR pumping at frequencies: a) 2.6 GHz, b) 9.0 GHz, T=300K are shown by symbols with fitting functions given by solid blue lines. Dashed lines show contributions of $V_Q$, $V_{AMR}^S$ and $V_{AMR}^A$, marked by appropriate subscripts.

Induced by FMR pumping spin current across a metal/ferromagnet interface $j_s$ is determined by the changing of magnetization and the components proportional to the real ($\operatorname{Re} g^{\uparrow\downarrow}$) and imaginary ($\operatorname{Im} g^{\uparrow\downarrow}$) parts of spin mixing conductance: [6, 40, 41]:

$$j_S = \frac{h}{4\pi}\left(\operatorname{Re} g^{\uparrow\downarrow}\, m\frac{dm}{dt} + \operatorname{Im} g^{\uparrow\downarrow}\, \frac{dm}{dt}\right) \quad (2)$$

where m is normalized magnetization in ferromagnetic. A family of magnetic field dependences of the transmitted microwave coefficient $S_{12}(H)$ recorded under FMR pumping at fixed frequency within band F= 2-20 GHz were used for determination of $\operatorname{Re} g^{\uparrow\downarrow}$ and $\operatorname{Im} g^{\uparrow\downarrow}$ (see Appendix and Ref. [42]).

The Gilbert spin damping $\alpha$ characterizes by independent on FMR frequency spin precession attenuation. A broadening of the width $\Delta H$ is seen in increase of $\alpha$. In the normal metal/ferromaget heterostructure an observation of increase in $\alpha$ caused by spin current generation across the interface was discussed e.g. in [6, 21, 40, 41]. Figure 4a shows the linewidth dependences obtained from the spectra $S_{12}(H)$ for both the LSMO film and the SIO/LSMO heterostructure as a function of the microwave frequency. The Gilbert spin damping $\alpha$ and the linewidth broadening $\Delta H_0$ can be determined using equation [41]:

$$\Delta H(F) = 4\pi\alpha F/\gamma + \Delta H_0 \quad (3)$$

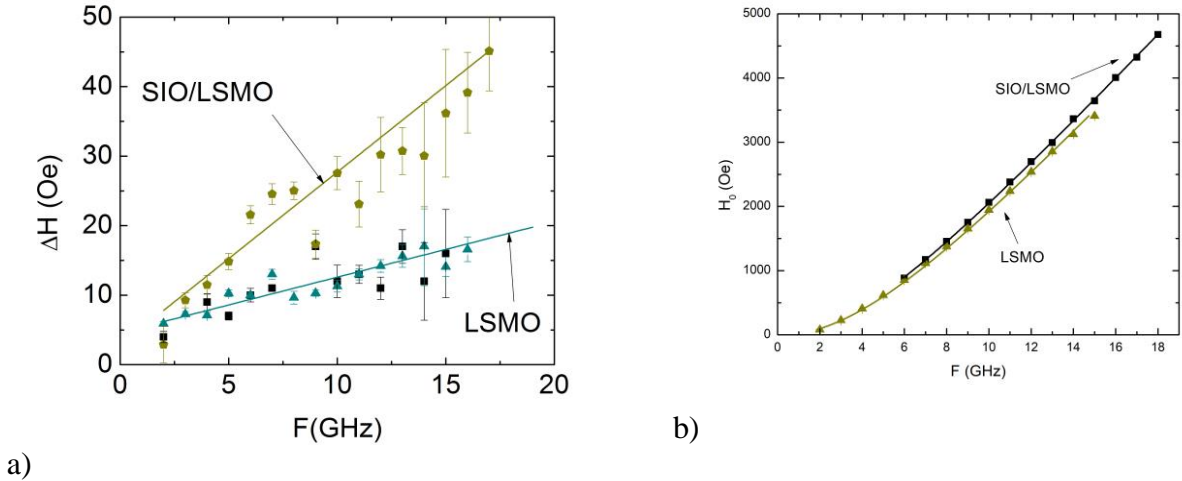

a) b)

Fig. 4.a) Frequency dependence of the FMR linewidth $\Delta H$ for the LSMO film and the SIO/LSMO heterostructure. The solid lines show linear approximations of the experimental data for $\Delta H(F)$ eq. (3). b) Frequency resonance field dependence $H_0(F)$. The solid lines show approximations (6) of the experimental data for H(F) for M=370 G, $H_u$ =11 Oe LSMO film and SIO/LSMO heterostructure with the same M and $H_u$ but change gyromagnetic ratio $\gamma$,.

In this case we neglect other sources of spin damping (see, for example, [43]). Spin damping for LSMO film is $\alpha_{LSMO}=2.0\pm0.2\ 10^{-4}$, and after SIO film sputtering on top of LSMO in SIO/LSMO heterostructure it increases up to $\alpha_{SIO/LSMO}=6.7\pm0.8\ 10^{-4}$. The frequency independent broadening of resonant linewidth $\Delta H_0$ at low frequencies F<6 GHz $\Delta H_0=6\pm1$ Oe is small and could be attributed to a magnetic inhomogeneity. At higher frequencies broadening of FMR linewidth and an increase of Gilbert damping allows to estimate the real part of spin mixing conductance as follows [6,40,41,46]:

$$\text{Re}\,g^{\uparrow\downarrow} = \frac{4\pi M d_{LSMO}}{g\mu_B}(\alpha_{SIO/LSMO}-\alpha_{LSMO}) \qquad (4),$$

where LSMO film magnetization M=370 G, $d_{LSMO}$=30 nm is LSMO film thickness, $\mu_B=9.274\cdot10^{-21}$ erg/G is the Bohr magnetron, Lande factor g=2. From $S_{12}(H)$ data from FMR pumping at F=9 GHz we got $\text{Re}\,g^{\uparrow\downarrow}=(3.5\pm0.5)\cdot10^{18}$ m$^{-2}$. Note, $\text{Re}\,g^{\uparrow\downarrow}=1.3\ 10^{18}$ m$^{-2}$ was obtained in [44] for the SIO/LSMO heterostructure fabricated by laser ablation. According to [45], when the SIO film thickness changes from 1.5 to 12 nm $\text{Re}\,g^{\uparrow\downarrow}$ for SIO/LSMO heterostructure changes from 0.5 to 3.6 $10^{19}$ m$^{-2}$ respectively.

Theory based on the spin-exchange interaction between localized moments and conductivity electrons shows that the deterministic material properties for $\text{Re}\,g^{\uparrow\downarrow}$ are the electrical resistivity $\rho_{SIO}$ and the spin diffusion length $\lambda_{SIO}$ of N metal [40, 46, 47]:

$$\text{Re}\,g^{\uparrow\downarrow} \approx (h/e^2)/(\rho_{SIO}\lambda_{SIO}). \qquad (5)$$

Here $h/e^2 \approx 25.8\ k\Omega$ is the quantum of resistance. Equation (5) is valid for a transparent interface and represents the lowest limit of $\text{Re}\,g^{\uparrow\downarrow}$. For parameters $\lambda_{SIO}$ = 1 nm [44] and $\rho_{SIO}$ = 3 $10^{-4}\ \Omega$ cm [36] we got $\text{Re}\,g^{\uparrow\downarrow}\approx 8.6 \times 10^{18}$ m$^{-2}$. The estimation roughly agrees with the experimental data [48]. Nevertheless, Eq. (5) gives just a qualitative insight for the impact of material parameters on $\text{Re}\,g^{\uparrow\downarrow}$ and doesn't take into account SOI and influence of magnetic inhomogeneities on interface properties.

Fig. 4b shows the dependences of the resonant field $H_0$ on the microwave frequency F for the LSMO film and SIO/LSMO heterostructure when the magnetic field is directed along the easy magnetization axis. It is possible to determine the magnetization M and uniaxial magnetic anisotropy $H_u$ for LSMO film from fitting the experimental curves $H_0(F)$ by Kittel relation:

$$F = \gamma(4\pi M + H_u + H_0)^{1/2}(H_0 + H_u)^{1/2}$$

(6)

In order to obtain the change in resonance filed $H_0$ caused by sputtering SIO over LSMO we performed fitting of $H_0(F)$ (6) dependences for SIO/LSMO heterostructure. As shown by measurements of the angular dependences for both LSMO films and SIO/LSMO heterostructures [36, 49] the variation of cubic anisotropy after SIO deposition can be neglected. Magnetization M as well the uniaxial magnetic anisotropy $H_u$ values differ slightly from those obtained from the $H_0(F)$ relation.. At high frequencies (above 10 GHz) a deviation of the $H_0(F)$ dependence of heterostructures from $H_0(F)$ for LSMO is observed (Fig. 4b). Taking fitting parameters for LSMO magnetization M=370 G, magnetic anisotropy $H_u$ =11 Oe and gyromagnetic relation γ=2.8 GHz/kOe for experimental data $H_0(F)$ we obtain a fitting curve which corresponds well to (6) given in Fig. 4b. At the same time for SIO/LSMO heterostructure $H_0(F)$ deviates. However, there is no physical grounds for changing M and $H_u$ after deposition of SIO on LSMO film. The observed deviation using approach in [40] can be fitted by a recalculated $H_0(f)$ function in terms of deviation in gyromagnetic ratio γ, which could be caused by presence of the imaginary component Im $g^{\uparrow\downarrow}$ in heterostructure . The relative change $\delta\gamma/\gamma_0$=0.036±0.001 caused by sputtering SIO on top of LSMO gives fitting curve in Fig.4b and allows to determine Im $g^{\uparrow\downarrow}$ [6, 40, 46]:

$$\delta\gamma/\gamma_0 = \text{Im} g^{\uparrow\downarrow} \frac{g_0 \mu}{4\pi M_S d_{LSMO}}$$

(7)

For the heterostructure shown in Fig. 4 we obtain (Im $g^{\uparrow\downarrow}$)$_{max}$= (46 ±1) $10^{19}$ m$^{-2}$. This value noticeably exceeds obtained earlier for platinum/ferromagnet structures [6, 46, 48]. Perhaps, the $H_u$ and M variations should be taken into account to obtain realistic $\delta\gamma/\gamma_0$. Again, as showed measurements of the angular dependences of the resonance field after SIO sputtering on top of LSMO there is a significant change in $H_u$ anisotropy. Fig.2Aa (see Appendix) shows the changes in $H_0(F)$ with increasing $H_u$. It can be seen that the theoretical dependences strongly deviate from the experiment with increasing $H_u$. On the other hand with decreasing M from 370 G to 330 G the obtained dependence (6) describes well the data for SIO/LSMO heterostructure (see Fig.A2b, Appendix). At the same time, the value $\delta\gamma/\gamma_0$ lays within the measurement error $\delta\gamma/\gamma_0 \approx 0.001$, which gives an estimate below (Im $g^{\uparrow\downarrow}$)$_{min}\approx 10^{19}$ m$^{-2}$. As shown in [6, 40, 45, 46] for Im $g^{\uparrow\downarrow}$ comparing with Re $g^{\uparrow\downarrow}$ the properties of the interface between the ferromagnetic and the normal metal and the quality of the interface may play an important role. The measurements of the spin Hall magnetoresistance for demonstrate in Im $g^{\uparrow\downarrow}$ 3 times and 10 times larger than Re $g^{\uparrow\downarrow}$ in Pt/EuS and

W/EuO heterostructures, correspondingly [50, 51]. Note, in addition to a possible change of in-plane magnetization of LSMO an appearance of out-of-plane magnetic moment in the direction of normal to the interface could also takes place as has been observed in manganite/iridate superlattices [29].

### 4. Spin magnetoresistance

The measurement of magnetoresistance (MR) is shown to be an effective method for probing the spin related properties in the metal layer such as the spin Hall angle and the spin diffusion length [7-12, 50, 51]. If the SHE and ISHE processes are coupled by SOI a change in MR becomes spin dependent [3, 8]. The relationship of charge current density $\vec{j}_Q$ induced by ISHE and spin current $j_S$ characterized by spin Hall angle $\theta_{SH}$ is determined by the following equation [4, 41]:

$$\vec{j}_Q = \theta_{SH} \frac{2e}{\hbar} \left[ \vec{n} \times \vec{j}_S \right] \tag{8}$$

where $\vec{n}$ is unit vector of spin momentum direction.

The heterostructure SIO/LSMO deposited on NGO substrate was fabricated for measurements in planar Hall configuration with electric contacts at the edges (Fig.5). Either a voltage $V_L$, proportional to longitudinal MR, or a transverse voltage $V_T$ (transverse MR) were measured using low noise frequency selective lock-in amplifier when current I=0.5 mA at frequency F~ 1.1 kHz was applied along the x-direction. External magnetic field H was swept in sequence: $0 \rightarrow H_+ \rightarrow 0 \rightarrow H_- \rightarrow 0$ with a step $\Delta H = H_{MAX}/N$, N=200÷1500, $H_+ = H_{MAX}$, $H_- = -H_{MAX}$. For changing the angle φ between magnetic field and current I the substrate was rotated relative to the direction of field H set by the Helmholtz coils. Note, besides of appearance of SMR the anisotropic MR (AMR) also contributes to MR response.

Comparing the MR for SIO/LSMO heterostructure with a single LSMO film, for which only AMR was a priori anticipated, the impact of SMR was revealed. The magnetic field dependences for normalized MR for LSMO and SIO/LSMO are given in Fig.6. Resistance $R_0$ at H=0 was used for normalization.

As seen from Fig.6a for LSMO film the 90 degree rotation of φ to the sign change of MR demonstrating a linear rise with H at $0<H<|H_{MAX}|$. An oscillating behavior of MR as cos2φ between curves (1) and (2) was observed as expected for AMR. This is seen also in Fig.6b, where H-φ plane is given for LSMO with MR given by a colored scale. Similar dependencies for SIO/LSMO are presented in Fig.6c,d. In this case MR R(H) lays always below $R_0$ and is negative.

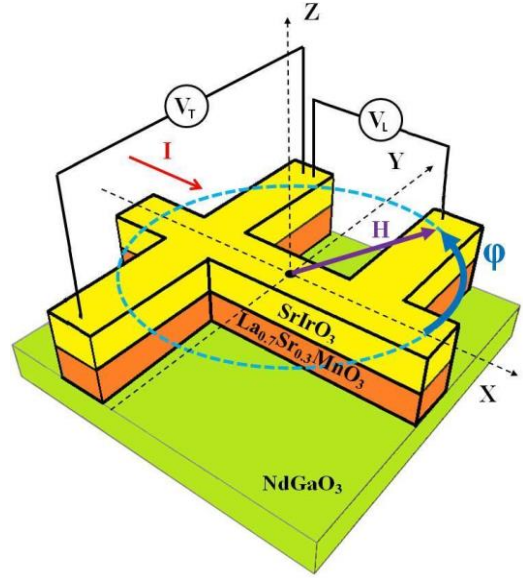

Fig.5. SIO/LSMO heterostructure on (110) NGO substrate and a 4-probe MR measurement scheme used for transverse $V_T$ and longitudinal $V_L$ voltage output terminals. Angle φ between direction of H field and current I was changed by rotation of substrate in X-Y plane.

Fig.7 demonstrates the longitudinal responses of oscillating angular dependences $\Delta R/R_0$ for LSMO film and for SIO/LSMO heterostructure. The first one is associated with the AMR, while the second contains contributions of both the SMR and AMR. The AMR is observed in metallic ferromagnets and shows an oscillating dependence of electric resistance on the angle φ between the direction of electric current and the magnetization. For ferromagnets with weak anisotropy the relation for longitudinal MR is simplified and the angle φ could be counted between I current direction and in-plane magnetic field H (see for example [9, 52])

$$\left(\frac{\Delta R}{R_0}\right)_L = \left(\frac{\Delta R}{R_0}\right)_{AMR} \cos^2\varphi = \left(\frac{\Delta R}{2R_0}\right)_{AMR}(1+\cos 2\varphi) \tag{9}$$

In Fig.7 curve (1) shows dependence (9) for MR of LSMO film. The phase shift ϕ☐ in cos2φ dependence was used to take into account that the easy axis of the LSMO was shifted from the edge of the substrate, taken as the point of the substrate rotation, φ=0. Note, in normal metals (without magnetic order) electrons with spin up and spin down degenerate and AMR is absent. As mentioned earlier, in a bilayer structure with ferromagnet and nonmagnetic metal exhibiting strong SOI the longitudinal MR contains SMR. In such structures a charge current generates pure spin current [7, 53] with an efficiency character-ized by spin Hall angle $\theta_{SH}$. For SIO/LSMO heterostructure the longitudinal SMR takes the form containing a similar to AMR component [9, 52]:

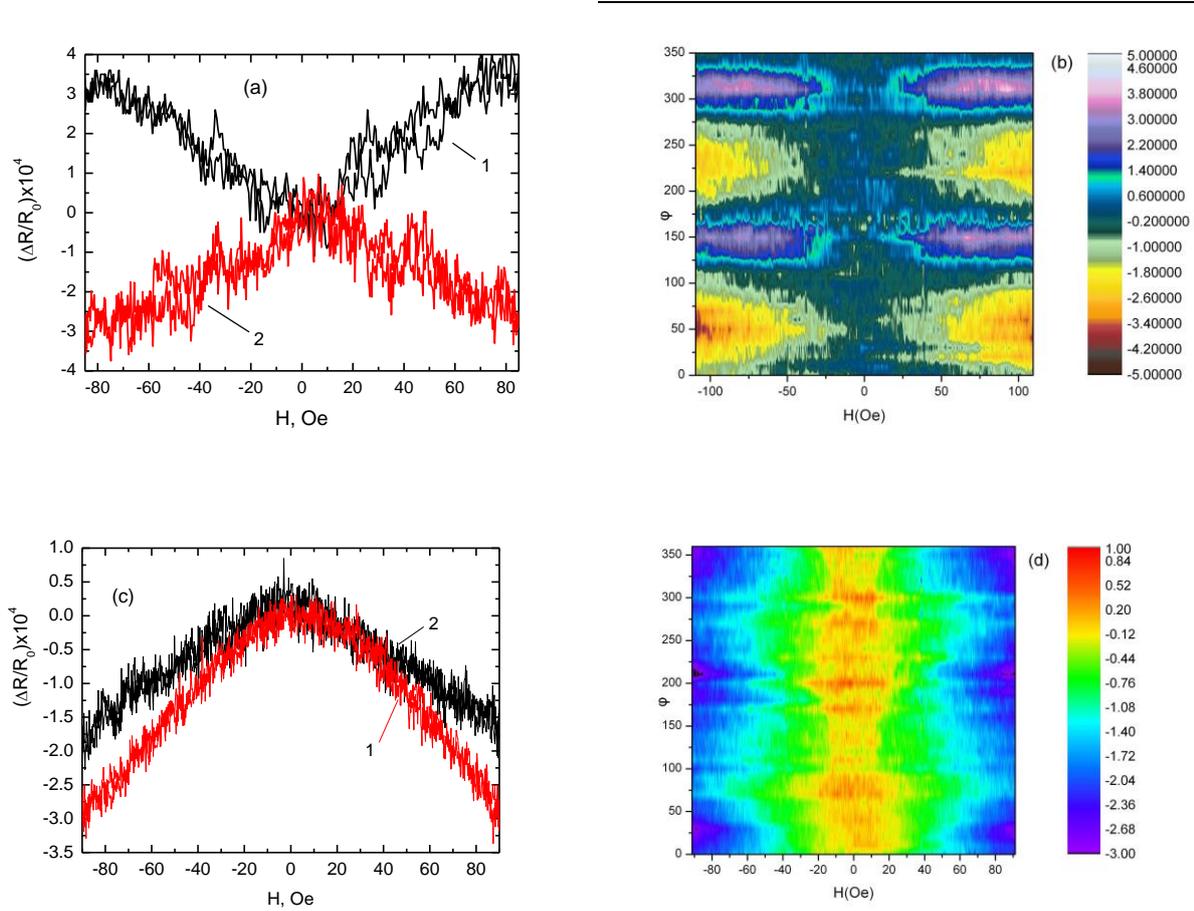

Fig.6. Magnetoresistance ($\Delta R/R_0$), normalized on $R_0=R(H=0)$: (a) magnetic field dependence for LSMO film at two angles $\varphi=140°$ (1) and $\varphi=230°$ (2) and (b) H-$\varphi$ cut of 3D magnetoresistance image; (c) SIO/LSMO at $\varphi= 30°$ (1) and $\varphi= 80°$ (2). (d) H-$\varphi$ image. The ($\Delta R/R_0$) amplitudes (b, d) are given in colored scale multiplied by $10^4$.

$$\left(\frac{\Delta R}{R_0}\right)_L = \left(\frac{\Delta R}{R_0}\right)_0 + \left(\frac{\Delta R}{2R_0}\right)_1 (1+\cos 2\varphi) \tag{10}$$

where

$$\left(\frac{\Delta R}{R_0}\right)_0 = -\theta_{SH}^2 \frac{2\lambda_{SIO}}{d_{SIO}} \tag{11}$$

$$\left(\frac{\Delta R}{R_0}\right)_1 = \theta_{SH}^2 \frac{\lambda_{SIO}}{d_{SIO}} \mathrm{Re} \frac{2\lambda_{SIO}\rho_{SIO}(\mathrm{Re}G^{\uparrow\downarrow} + i\,\mathrm{Im}G^{\uparrow\downarrow})}{1+2\lambda_{SIO}\rho_{SIO}(\mathrm{Re}G^{\uparrow\downarrow} + i\mathrm{Im}G^{\uparrow\downarrow})} \tag{12}$$

In (12) the imaginary part of spin mixing conductance Im $G^{\uparrow\downarrow}$ is considered as well. Taking resistivity of the SIO film $\rho_{SIO}$ = 3 $10^{-4}$ $\Omega$ cm) [36] and parameters obtained in part 3: Re $G^{\uparrow\downarrow}$ = 1.35 $10^{10}$ cm$^{-2}$ $\Omega^{-1}$ and (Im $G^{\uparrow\downarrow}$)$_{min}$= 3.88 $10^{10}$ cm$^{-2}$ $\Omega^{-1}$, $2\lambda_{SIO} \rho_{SIO}$ Re $G^{\uparrow\downarrow}$ =0.81 and $2\lambda_{SIO} \rho_{SIO}$ Im$G^{\uparrow\downarrow}$=2.33. For $\lambda_{SIO}$ = 1 nm [44] and d$_{SIO}$=10 nm and ($\Delta R/R_0$)$_0$ =-(2.2 ±0.05) $10^{-4}$ we got $\theta_{SH}$ =0.033±0.005 For ($\Delta R/R_0$)$_1$ = (0.7±0.14) $10^{-4}$ we obtain $\theta_{SH}$ =0.03±0.01.

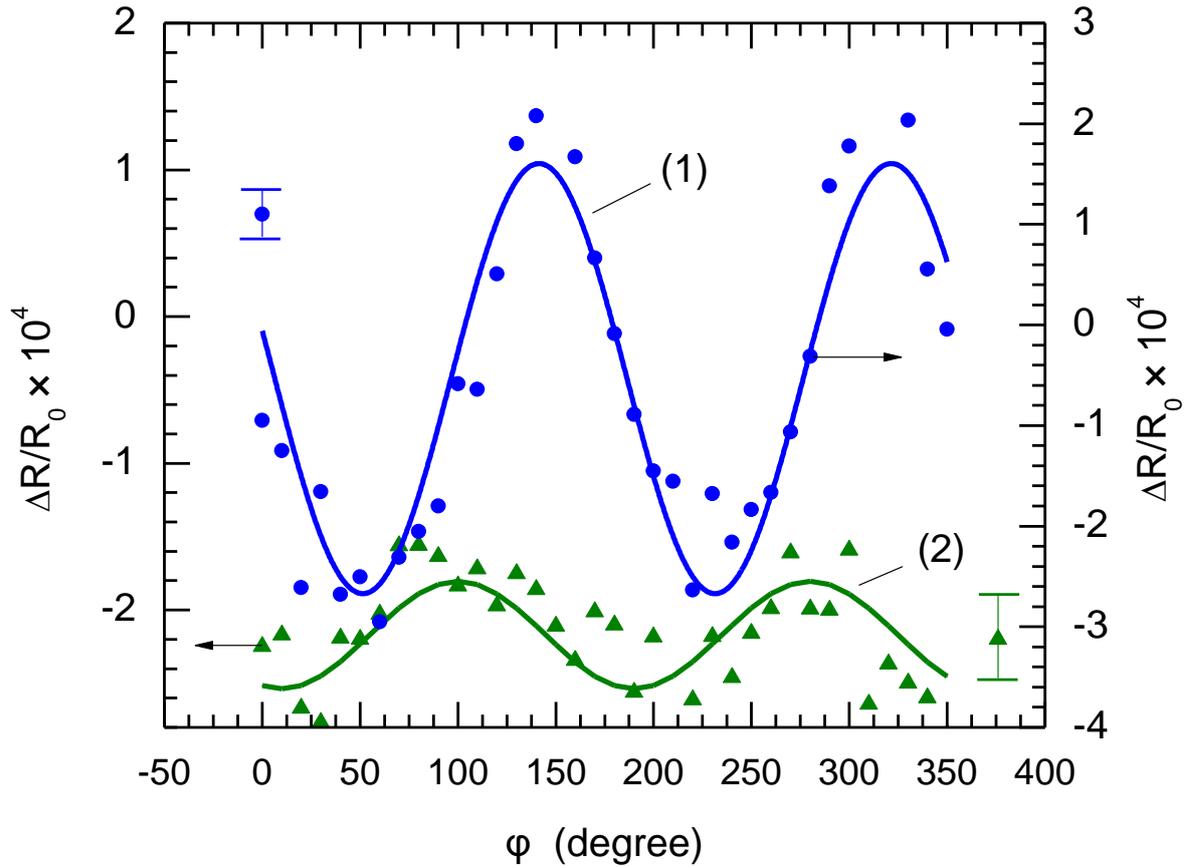

Fig.7. Angular dependencies of longitudinal MR for LSMO film (filled circles) and SIO/LSMO heterostructure (triangles) at H$_{MAX}$=90 Oe. Fitting dependences ~ cos(2$\varphi$+$\phi$) functions with $\phi$ shift are shown by solid lines (1) LSMO film, (2) SIO/LSMO heterostructure. An uncertainty is shown by an error bar.

In calculation of SMR, the data were taken from measurements at magnetic fields smaller than the H$_S$ saturation field (see Fig. 3A in Appendix ). As can be seen from Fig. 3A for LSMO film and SIO/LSMO heterostructure a difference of H$_S$ fields does not exceed 20%. The calculated SMR value increases with ($\Delta R/R_0$) almost linearly with the magnetic field up to H=H$_S$ and then at H>>H$_S$ saturates where Hanle effect takes place [559].

A possible reason for the strong decrease in the measured SMR of the heterostructure is the shunting of the LSMO and SIO films resistances by the conductive layer at the SIO/LSMO interface [36]. In this case the resistance ($R_H^1$) of the SIO/LSMO heterostructure can be modeled as a parallel connection of resistance of the upper layer of SIO film $R_S$ and resistance of the bottom LSMO layer $R_L$. $R_H^1 = R_S R_L/(R_S+R_L)$ and resulted in measured resistance of heterostructure ($R_H$) becomes smaller than calculated $R_H^1$ [36]. A possible solution of the problem is to account a parallel connection of the interface resistance $R_I$. Using sheet resistance of interface SIO/LSMO $R_I = \rho_I/d_S$ we get $\rho_I = 8\cdot 10^{-6}$ Ω·cm supposing the thickness of interface is of order of 1 nm [36]. A small resistivity of the interface may indicate an existence of a thin well conducting layer as a 2D electron gas with high mobility, possibly as at interfaces [56, 57].

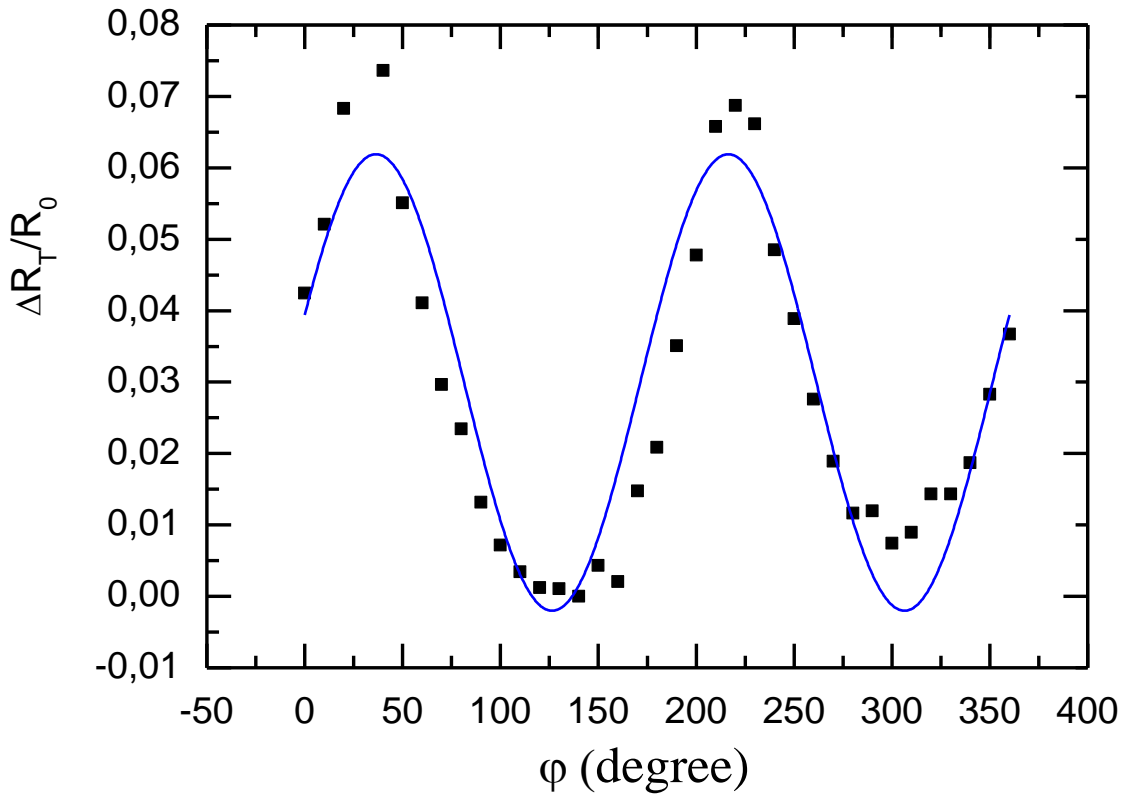

Fig.8. Angular dependencies of transverse MR for SIO/LSMO heterostructure (filled squares) at $H_{MAX}$=90 Oe. Fitting dependences ~ $\sin(2\varphi+\phi)$ functions are shown by solid lines.

The model [58] for iridate/manganite interface show that charge transfer at the interface from the half-filled spin-orbit entangled $J_{eff} = 1/2$ state to the empty $e^{\uparrow g}$ states may occur. The charge leakage from iridate makes it hole doped, while the manganite side of SIO/LSMO heterostructure becomes electron doped. The doped carriers make both sides metallic. La and Sr doping across the interface or oxygen doping that give additional conductivity are not excluded [29,59].

In our measurements configuration (see Fig.6) with current direction along x, for voltage along y we get [9,52] the transverse MR, which also referred as a planar Hall effect:

$$\left(\frac{\Delta R}{R}\right)_T = \left(\frac{\Delta R}{2R}\right)_1 \sin 2\varphi + \left(\frac{\Delta R}{R}\right)_2 \cos\theta \quad (13)$$

where $\theta$ is the angle for out-of-plane magnetization along z (not shown in Fig.6) relative charge current I. The measured SMR value is $(\Delta R/R)_1 = 0.032\pm0.002$ which gives $\theta_{SH} = 0.65\pm0.05$, considering that in planar Hall effect configuration the conductive layer at the interface SIO/LSMO has no influence. So, as in (13) we have the first term of resistance change in amplitude with angle $2\varphi$. Note, an additional term may arise from the magnetization directed perpendicular to the plane [59], determined by the angle between the magnetization and the substrate plane.

## 5. Conclusion

The transmission electron microscope investigation and X-ray diffraction measurements of $SrIrO_3/La_{0.7}Sr_{0.3}MnO_3$ heterostructure show epitaxial growth of both films in heterostructure with smooth interface. It was shown that in regime of ferromagnetic resonance the voltage response induced by anisotropic magnetic resistance is compared with the response induced by generation of spin current flowing across interface. The real and imaginary parts of spin mixing conductance of heterostructure were determined from frequency dependence of FMR magnetic field. Obtained data for real part of spin mixing conductance agrees with the experimental data obtained previously and gives a realistic qualitative insight into impact of material parameters used in experiment. The imaginary part of spin mixing conductance of $SrIrO_3/La_{0.7}Sr_{0.3}MnO_3$ interface was found extremely high. It depends on the properties of the interface between the ferromagnetic and the normal layer with strong spin-orbit coupling. The spin Hall angle was determined by measuring the spin magnetoresistance. An influence of anisotropic magnetoresistance on measured data of spin Hall magnetoresistance was observed. The interlayer with high conductivity at the interface of $SrIrO_3/La_{0.7}Sr_{0.3}MnO_3$ heterostructure shunts measured resistance in longitudinal mode, but does not in transverse MR measurements. Estimation of spin Hall angle for the interface turns out to be higher than for the case of transverse MR in the interface with Pt film.

The authors are grateful to V.V. Demidov, A.M. Petrzhik, K.L. Stankevich, T.A. Shaikhulov for their help and discussion of experimental results.

This work was carried out within the framework of the state task of the Kotel'nikov Institute of Radio Engineering and Electronics Russian Academy of Sciences. The study was carried out using the Unique Science Unit "Cryointegral" (USU #352529), which was supported by the Ministry of Science and Higher Education of Russia (Project No. 075-15-2021-667). Structural investigations were carried out using the equipment of the Center "Material Science and Metallurgy" with the financial support of Ministry of Education and Science of Russian Federation (Agreement № 075-15-2021-696).

Appendix

The typical $S_{12}(H)$ spectrum is presented in Fig. 1A . It is approximated using sum of the Lorentz line (L) for the imaginary part of FMR and the dispersion relation for the real part (D)[42]. From fitting the experimental curves with these two components it is possible to determine the resonance field ($H_0$) and the width of FMR line ($\Delta H$).

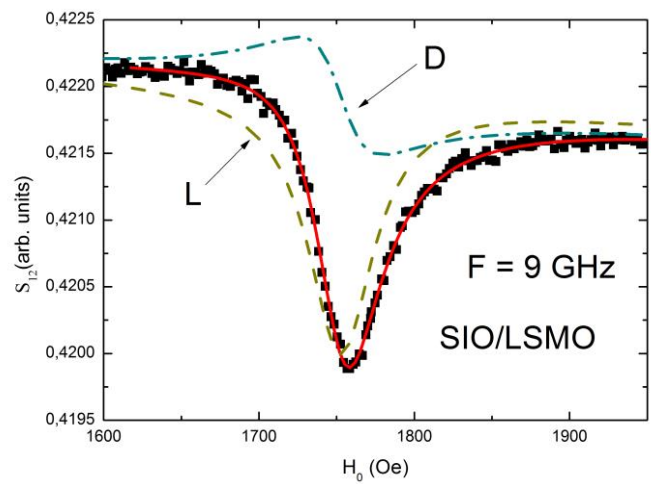

Fig.1A. The magnetic filed dependence $S_{12}$ (H) for SIO/LSMO heterostructure under microwave radiation at f= 9 GHz in microstrip configuration. The solid line shows the approximation of the spectrum, L is Lorentz line and D is dispersion relation.

As showed measurements of the angular dependences of the resonance field, after SIO sputtering on the top of LSMO there is a change in $H_u$ anisotropy field. Figure 2A.a) shows the changes in $H_0(F)$ with increasing $H_u$. It can be seen that the theoretical dependences strongly deviate from the experiment with increasing $H_u$ from 11 to 100 Oe. On the other hand, with decreasing M from 370 G to 330 G the obtained dependence (eq. 6) describes well the data for SIO/LSMO (see Fig.2A.b).

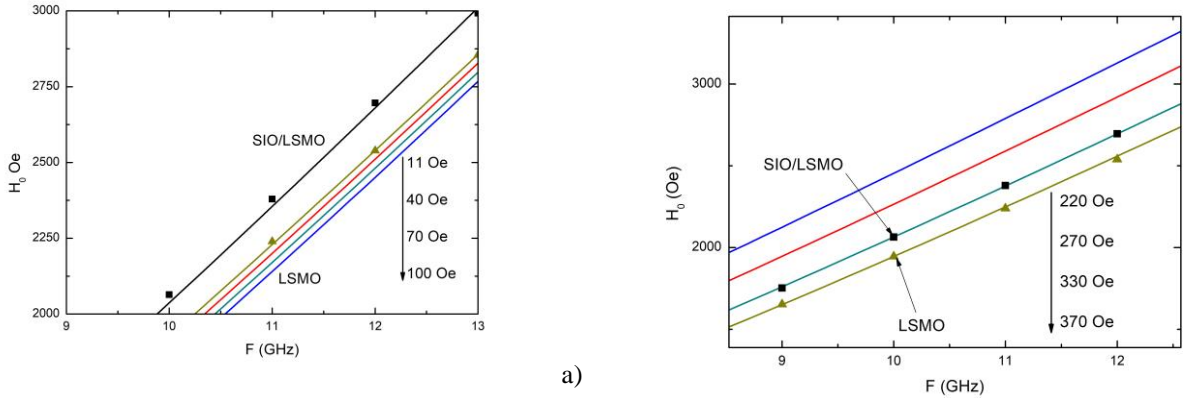

Fig 2A. a) Calculated by eq. (5) $H_0(F)$ dependences for $H_u$ = 11, 40, 70, 100 Oe. The curves arranged from top to bottom as shown by the arrow, squares and triangles denote experimental values for SIO/LSMO and LSMO, respectively. (b) $H_0(F)$ for 4 values $M_0$ = 220, 270, 330, 370 G (see arrow from top to bottom), squares and triangles denote the experimental values for SIO/LSMO and LSMO, respectively.

Fig.3A. shows magnetic field dependences of the normalized magnetization of the SIO/LSMO heterostructure and LSMO film measured using the Kerr magneto-optical effect. In calculation of SMR, the data were taken from measurements at magnetic fields smaller than the $H_S$ saturation field. As can be seen from Fig. 3A b) for LSMO film and a) SIO/LSMO heterostructure a difference of $H_S$ fields does not exceed 20%. SMR value increases with ($\Delta R/R_0$) almost linearly with the magnetic field up to H=$H_S$ and then at H>>$H_S$ saturates.

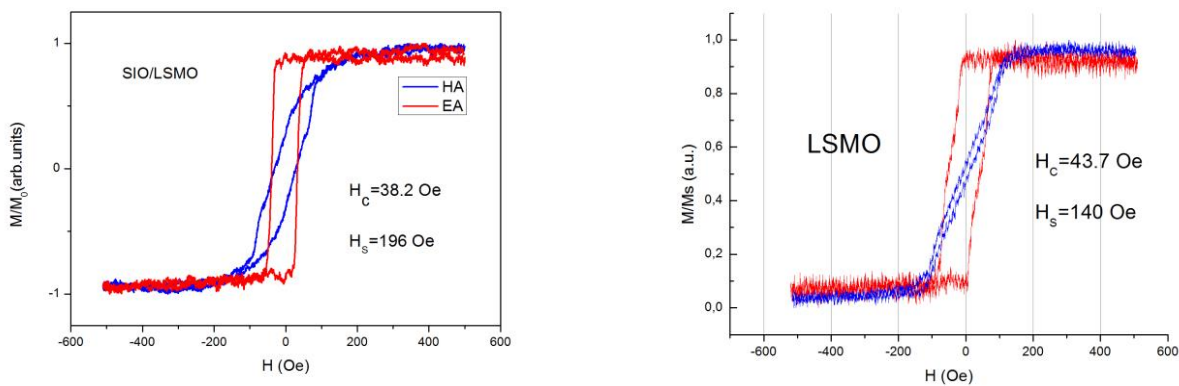

a)                                                                  b)

Fig.3A. Magnetic field dependence of the normalized magnetization of the SIO/LSMO heterostructure (a) and LSMO film (b), measured using the Kerr magneto-optical effect. Dependences EA (red line ) correspond to easy axis, HA (blue lines) – hard axis.